# SELF-MEDRAG: A SELF-REFLECTIVE HYBRID RETRIEVAL-AUGMENTED GENERATION FRAMEWORK FOR RELIABLE MEDICAL QUESTION ANSWERING

JESSICA RYAN[1,*], ALEXANDER I. GUMILANG[1], ROBERT WILIAM[1],
DERWIN SUHARTONO[1]

[1]School of Computer Science, Bina Nusantara University, Anggrek Campus,
Jl. Raya Kb. Jeruk No. 27, 11530, Jakarta, Indonesia
*Corresponding Author: jessica.ryan@binus.ac.id

**Abstract**

Large Language Models (LLMs) have demonstrated significant potential in medical Question Answering (QA), yet they remain prone to hallucinations and ungrounded reasoning, limiting their reliability in high-stakes clinical scenarios. While Retrieval-Augmented Generation (RAG) mitigates these issues by incorporating external knowledge, conventional single-shot retrieval often fails to resolve complex biomedical queries requiring multi-step inference. To address this, we propose Self-MedRAG, a self-reflective hybrid framework designed to mimic the iterative hypothesis-verification process of clinical reasoning. Self-MedRAG integrates a hybrid retrieval strategy, combining sparse (BM25) and dense (Contriever) retrievers via Reciprocal Rank Fusion (RRF) to maximize evidence coverage. It employs a generator to produce answers with supporting rationales, which are then assessed by a lightweight self-reflection module using Natural Language Inference (NLI) or LLM-based verification. If the rationale lacks sufficient evidentiary support, the system autonomously reformulates the query and iterates to refine the context. We evaluated Self-MedRAG on the MedQA and PubMedQA benchmarks. The results demonstrate that our hybrid retrieval approach significantly outperforms single-retriever baselines. Furthermore, the inclusion of the self-reflective loop yielded substantial gains, increasing accuracy on MedQA from 80.00% to 83.33% and on PubMedQA from 69.10% to 79.82%. These findings confirm that integrating hybrid retrieval with iterative, evidence-based self-reflection effectively reduces unsupported claims and enhances the clinical reliability of LLM-based systems.

## 1. Introduction

Large Language Models (LLMs) have demonstrated strong potential for medical question answering (QA) by synthesizing complex biomedical knowledge and supporting both clinicians and patients [1,2]. However, despite their strengths, LLMs remain limited by hallucinations, overconfident reasoning, and dependence on static pre-training corpora that cannot keep pace with rapidly evolving medical evidence. In high-stakes clinical scenarios, these shortcomings pose serious risks [3].

Retrieval-Augmented Generation (RAG) has emerged as a promising solution by grounding LLM outputs in external evidence [4]. Yet, conventional RAG typically follows a single-shot pipeline: retrieve once, answer once. While effective in general domains, this static workflow often fails in medical contexts where queries require iterative clarification, hypothesis revision. and progressive evidence gathering processes that closely mirror how clinicians' reason through differential diagnoses. A single retrieval pass is frequently insufficient for capturing nuanced biomedical knowledge or resolving ambiguous questions [5,6].

Recent research addresses these limitations through self-reflective and iterative RAG methods. Self-RAG enables models to critique their own answers and request additional evidence, but their full implementation is complex, computationally heavy, and not specifically optimized for the domain of clinical reasoning [7]. Conversely, lightweight iterative RAG systems perform multiple reasoning-retrieval cycles but often lack domain-aware mechanisms for evaluating whether retrieved evidence truly supports the model's rationale. As a result, these approaches either introduce unnecessary complexity or fail to capture the domain-aware evidence requirements of medical QA [8,9].

To address the challenge, we propose **Self-MedRAG**, a self-reflective hybrid RAG framework for reliable medical QA. Self-MedRAG mimics the stepwise reasoning process of clinicians by combining hybrid retrieval, a lightweight self-reflection module using existing NLI and LLM models, and iterative query refinement. This workflow enables the framework to progressively strengthen factual grounding while maintaining clinical coherence and transparency.

Our contributions are:

1. We present **Self-MedRAG**, an iterative RAG framework that integrates retrieval, generation, verification, and query refinement into a unified loop tailored for medical QA.
2. We implement a lightweight self-reflection mechanism using off-the-shelf NLI and LLM models to guide additional retrieval and enhance evidence alignment without heavy computation.
3. We empirically demonstrate that Self-MedRAG improves factual accuracy, evidence grounding, and clinical safety on multiple medical QA benchmarks.

## 2. Related Works

### 2.1. Medical Question Answering

Medical question answering (QA) benchmarks evaluate a model's ability to generate clinically reliable, evidence-grounded responses. Widely used datasets include MedQA for diagnostic reasoning from medical exams, PubMedQA for evidence-based biomedical inference over research abstracts, and MMLU-Medical and MedMCQA for broad medical knowledge testing [10–13]. The MIRAGE benchmark further unifies these resources by providing a standardized RAG evaluation setup, pairing questions from multiple medical QA with curated retrieval corpora to assess grounding quality, retrieval dependence, and factual accuracy, advancing retrieval-augmented methods in medical QA [14].

### 2.2. Large Language Models in Medical Domain

With the emergence of large-scale LLMs, many studies explored applying pretrained language models to medical QA. Models such as BioBERT, PubMedBERT, Med-PaLM, and PMC-LLaMA have demonstrated strong performance on biomedical by reasoning over unstructured clinical text and handling complex multi-step questions [15–17]. However, despite these advances, LLMs still face significant limitations in clinical contexts, including hallucinations, reliance on static or incomplete pretraining data, sensitivity to demographic and domain shifts, and challenges related to bias, explainability, and equitable decision-making, which limits reliability for clinical decision support [18]. Prior work also evaluated API models like GPT-4.1 and DeepSeek, along with open-source LLMs, under RAG and SFT, providing baselines for hybrid retrieval and evidence-grounded medical QA [19].

### 2.3. Retrieval-Augmented Generation and Knowledge Graphs

Retrieval-Augmented Generation (RAG) improves the reliability of LLMs by grounding their outputs in external evidence. Classical RAG systems rely on a single retrieval step using sparse methods such as BM25, which match queries and documents through keyword-based term overlap, or dense models such as E5, BGE, and Contriever, which encode text into semantic vector representations to capture deeper meaning beyond exact wording [20–22]. Hybrid retrieval strengthens coverage by combining outputs from multiple retrievers through methods like Reciprocal Rank Fusion (RRF) or Weighted Fusion, then applying a cross-encoder reranker to identify the most relevant passages for the LLM [23,24]. In medical settings, domain-adapted retrievers like BioBERT, PubMedBERT, and MedCPT improve precision by leveraging biomedical corpora and PubMed search logs, forming the domain-specific backbone of medical RAG systems [25,26].

Recent graph-based RAG systems extend document retrieval with knowledge graphs and biomedical ontologies, enabling models to access entity and relation-level information for structured reasoning over diseases, symptoms, and treatments. In the medical domain, frameworks like MedRAG and MedGraphRAG combine text retrieval with graph traversal or graph navigation to enhance evidence grounding and multi-hop inference. These approaches improve factuality and interpretability but introduce added complexity in graph construction, computational load, indexing, and maintenance [27–29].

### 2.4. RAG Workflows and Critic Models

RAG workflows have evolved from classical one-shot retrieval to advanced architectures that incorporate iterative retrieval, modular task decomposition, and critic-based refinement to progressively improve answer quality. Building on these advances, Agentic RAG further empowers the LLM to autonomously plan, decide when to retrieve or reformulate queries, verify evidence, and invoke external tools for complex multi-step reasoning [30–33]. Central to these advanced RAG paradigms is the critic module, responsible for ensuring factuality and context utility [34]. Examples include Self-RAG's learned reflection token for self-critique its own generated answer, identify unsupported claims, and decide whether additional retrieval is needed, and CRAG's (Corrective RAG) retrieval sufficiency evaluator for checking context relevance [7,35].

In the medical domain, several frameworks adapt these workflow ideas to clinical reasoning. MRD-RAG performs multiple rounds of retrieval and generation, simulating a clinician's diagnostic process, with each round building on previous answers and retrieved evidence to progressively refine conclusions [36]. i-Med-RAG adopts a lightweight iterative loop in the medical domain, progressively improving answer quality and retrieval coverage without heavy computational overhead [5]. SIM-RAG introduces a sufficiency critic that halts reasoning early when evidence is adequate, improving efficiency but relying on general-domain sufficiency signals that may not reflect clinical requirements [6]. These systems highlight the importance of iterative reasoning, critique, and evidence sufficiency in medical decision-support settings, but none fully integrate strong domain grounding with efficient self-reflection.

## 3. Methodology

### 3.1. Self-MedRAG Overview

Self-MedRAG is an iterative Retrieval-Augmented Generation framework for reliable medical question answering. As illustrated in Fig 1, the system follows a loop that mirrors how clinicians gather evidence, generate a provisional explanation, and reassess whether their reasoning is sufficiently supported. At each iteration, the model retrieves context passages, produces an answer and rationale, then evaluates that rationale through a scoring module. If parts of the rationale are unsupported or contradictory, the system reformulates the query to target missing information and repeats retrieval and generation. This reflection cycle allows Self-MedRAG to progressively strengthen factual grounding while ensuring that the final answer and rationale remain clinically coherent, and evidence based.

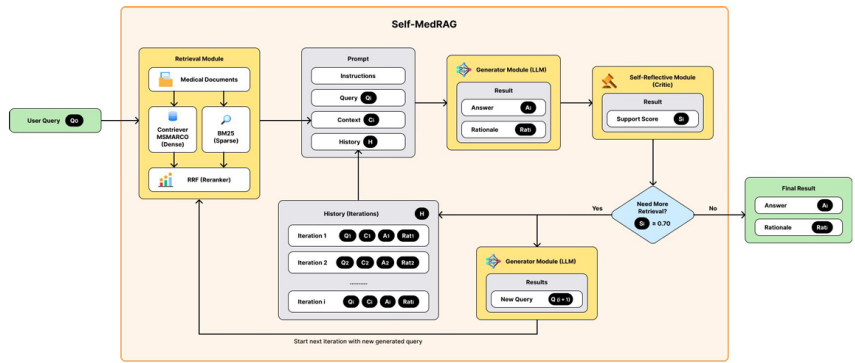

**Fig. 1. Self-MedRAG Pipeline**

The complete inference workflow of Self-MedRAG is summarized in Algorithm 1, which illustrates the iterative interaction among retrieval, generation, reflection, and refinement.

---
**Algorithm 1** Self-MedRAG Inference
---
**Require:** Initial question $Q_0$, History $H$, Retriever $R$, Generator $G$, Self-Reflection $SR$
**Ensure:** Final answer $A$ with rationale $Rat$
  $i \leftarrow 0$
  **while** true **do**
    **Input:** $Q_i$
    Retrieve context $C_i \leftarrow R(Q_i)$
    Generate answer and rationale $(A_i, Rat_i) \leftarrow G(Q_i, C_i, H)$
    Compute rationale support score $S_i \leftarrow SR(Rat_i, C_i)$
    **if** $S_i \geq 0.70$ **then**
      **return** $(A_i, Rat_i)$
    **else**
      Update history $H \leftarrow H \cup \{(Q_i, C_i, A_i, Rat_i)\}$
      Extract unsupported rationale $U_i \leftarrow \{r \in Rat_i \mid r \text{ unsupported by } C_i\}$
      Update query: $Q_{i+1} \leftarrow Q_i \cup U_i$
      $i \leftarrow i + 1$
    **end if**
  **end while**
---

### 3.2. Evaluation Datasets

We evaluate Self-MedRAG on a total of 1,000 questions randomly sampled from the test sets of two widely used medical question answering benchmarks: MedQA and PubMedQA. These datasets provide a rigorous testbed for assessing factual accuracy, evidence grounding, and clinical reasoning capabilities of our framework.

- **MedQA** is a multiple-choice medical exam dataset covering clinical scenarios from USMLE Step 1, Step 2, and Step 3 questions. Each instance contains a question, a set of answer options, and the correct answer [37].
- **PubMedQA** is a biomedical question answering dataset derived from PubMed abstracts. Each question is associated with a research abstract and a yes/no/maybe answer. We use the artificial subset of the test set [38] and removed any question that has maybe as its answer.

### 3.3. Retrieval Methods

The retrieval module provides the context used at each iteration of Self-MedRAG. For an input query $Q_i$, the system employs both sparse and dense retrieval pathways to maximize coverage of clinically relevant information.

BM25 is used as the sparse retriever, leveraging lexical matching and TF-IDF based scoring to capture high-precision biomedical terminology and exact phrase matches. To complement this, the system integrates Contriever-MSMARCO as the dense retriever, which encodes queries and documents into embedding vectors and ranks them through dot-product similarity. This enables retrieval of semantically related content that may not share explicit surface wording with the query. Dense embeddings are stored and searched efficiently through a FAISS index.

To combine the strengths of both retrieval signals, the top candidates from BM25 and Contriever are merged using Reciprocal Rank Fusion (RRF).

$$RRF(d) = \sum_{i=1}^{N} \frac{1}{K + rank_i(d)} \tag{1}$$

For each passage *d*, the fused RRF score is computed as shown on Eq. (1) where *K* = 60 follows standard practices. The resulting fused ranking yields the retrieval context $C_i$, which serves as the evidence foundation for the subsequent generation and verification steps.

### 3.4. Generator Module

The generator module is implemented using DeepSeek, a large language model (LLM) that serves as the primary reasoning and answer-generation component of Self-MedRAG. At each iteration *i*, the LLM receives a structured prompt containing: (1) the current query $Q_i$; (2) the retrieved context $C_i$; (3) system instructions encouraging evidence grounding, clinical caution, and avoidance of unsupported claims; and (4) the reasoning history *H*, when available, to maintain multi-step coherence.

DeepSeek then produces two outputs. It generates a task-formatted answer $A_i$, such as a binary choice or multi-choice selection depending on the question type, and a rationale $Rat_i$ that explains the decision and cites evidence from the retrieved context. This rationale is required to remain tightly linked to $C_i$, ensuring that the model's reasoning process stays transparent, traceable, and clinically grounded.

The reliability and iterative stability of the LLM enable it to refine its reasoning across cycles, integrate newly retrieved evidence, and correct earlier assumptions when necessary. The resulting answer-rationale pair *($A_i$, $Rat_i$)* is then forwarded to the Self-Reflective Module, which determines whether additional retrieval, query reformulation, or further refinement is required before the system finalizes the response.

### 3.5. Self-Reflective Module

The Self-Reflective Module evaluates whether the rationale $Rat_i$ generated by the LLM is adequately supported by the retrieved evidence context $C_i$. This verification determines whether the system can finalize the answer or must initiate another iteration. Each rationale statement is assessed for evidential support by comparing it against all passages in $C_i$. Two verification mechanisms are employed in the experiments:

- **NLI based Verification:** A RoBERTa-large-MNLI model is used to perform Natural Language Inference. For every evidence context and rationale pair, the evidence passage serves as the *premise* and the rationale statement as the *hypothesis*. The model outputs an entailment label and a confidence score. For each statement, the system selects the evidence passage with the highest entailment confidence as its best support.
- **LLM based Verification**: Llama~3.1-8B is prompted to behave as an NLI classifier, outputting *entailment*, *neutral*, or *contradict*. Its confidence scores are parsed from probability-style outputs in the same manner as the NLI model.

A rationale statement is marked as supported or unsupported based on whether its best entailment confidence exceeds the verification threshold τ = 0.5. The module then computes a rationale support score $S_i$, defined as the proportion of supported statements in $Rat_i$. If $S_i \geq \theta = 0.7$, the system returns $(A_i, Rat_i)$ as the final output. Otherwise, unsupported rationale elements are collected and used to construct a refined query $Q_{i+1}$, triggering an additional iteration. Thresholds τ and θ were selected based on preliminary validation experiments to balance retrieval depth and answer accuracy.

## 4. Results and Discussion

**Table 1. Comparison of Retrieval and Critic Configurations on PubMedQA and MedQA.**

| Method | Retrieval | Critic | PubMedQA | | MedQA | |
|---|---|---|---|---|---|---|
| | | | Acc/EM | F1 | Acc/Em | F1 |
| **Without Critic** | | | | | | |
| **Base RAG** | MedCPT | - | 65.97 | 62.60 | 39.90 | 40.49 |
| **Base RAG** | BM25 | - | 66.80 | 60.67 | 41.74 | 41.92 |
| **Base RAG** | Contriever | - | 67.90 | 64.41 | 43.30 | 41.15 |
| **Base RAG** | BM25 + Contriever + RRF | - | 69.10 | 64.45 | 80.00 | 79.93 |
| **With Critic** | | | | | | |
| **Self-MedRAG** | BM25 + Contriever + RRF | NLI (roberta-large-mnli) | 79.82 | 78.40 | 83.30 | 83.30 |
| **Self-MedRAG** | BM25 + Contriever + RRF | Llama3.1–8B | 78.76 | 77.31 | 82.90 | 82.90 |

The results presented in Table 1. demonstrate the performance trends across retrieval strategies and critic configurations. For Base RAG methods, hybrid retrieval using the combination of both BM25 and Contriever using Reciprocal Rank Fusion (RRF) achieves substantially stronger performance than any single retriever on both PubMedQA and MedQA dataset. While BM25 and Contriever individually reach accuracies of 66.80% and 67.90% on PubMedQA, their fusion through RRF slightly increases their performance accuracy to 69.10%. The effect is more pronounced on MedQA, where the method introduces a large jump of performance from 41.74% (BM25 alone) and 43.30% (Contriever alone) to 80.00% accuracy. This dramatic improvement proves that the fused retrieval using RRF provides broader coverage of clinically relevant evidence by integrating both high-precision lexical signals from BM25 and semantically aligned passages recovered by Contriever.

The lower performance observed with MedCPT retrieval (65.97% on PubMedQA and 39.90% on MedQA) shows the limitations of relying on a single embedding-based retriever. While MedCPT [39] is trained on biomedical corpora, its trained using contrastive learning objective which tends to produce a smoothed embedding space, which makes sentences collapse toward broad topical clusters [40]. This makes MedCPT effective at recovering general subject-matter passages but potentially less sensitive to fine-grained distinctions such as specific diagnostic criteria, reported effect sizes, or directionality of clinical outcomes.

In contrast to MedCPT, BM25 contributes precise lexical matching, and Contriever emphasizes semantic alignment through general-domain contrastive learning. When combined with Reciprocal Rank Fusion, these complementary signals results in more diverse evidence. This results in the context generated by the fused retrieval capturing both exact terminology and paraphrased biomedical reasoning, leading to a stronger performance.

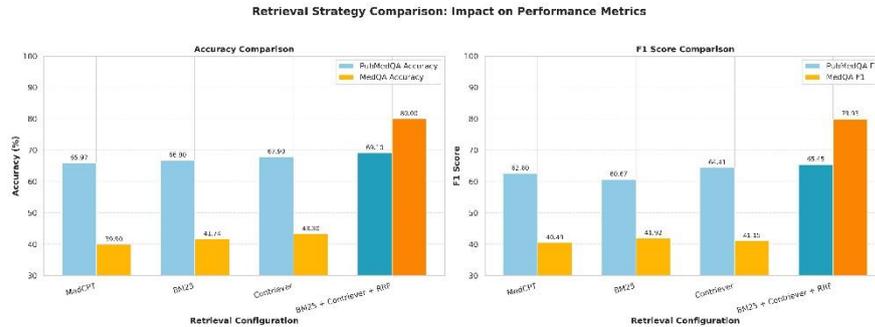

**Fig. 2. Comparison of Accuracy and F1 Score across different retrieval strategies on PubMedQA and MedQA datasets.**

To visually substantiate the performance gains of the hybrid retrieval approach, Fig 2 presents a comparative analysis of Accuracy and F1 scores across the PubMedQA and MedQA benchmarks. The bar plot highlights the limitations of relying solely on single retriever such as sparse (BM25) or dense (Contriever, MedCPT) retrievers, particularly on the MedQA dataset, where single-method performance of the method only hovers around 40-43% in both accuracy and F1 score. Notably, the integration of both BM25 and Contriever using Reciprocal Rank Fusion (RRF) bridged this performance gap, elevating MedQA accuracy to around 80%.

Beyond simple retrieval, our proposed Self-MedRAG framework achieved a substantial gain. As shown in Table 1, incorporating iterative verification improves the PubMedQA dataset accuracy performance from 69.10% (base RAG with with RRF) to 79.82%, the MedQA dataset accuracy performance increasing from 80.00% to 83.33% when using an NLI-based critic. These nearly ten-point gains on the PubMedQA and over three points on MedQA highlights the impact of using the Self-Reflective Module described in Section 3.5. Across iterations, the Generator Module produces answers and rationales that are refined based on updated retrieval contexts, allowing the system to correct unsupported assumptions and improve its reasoning.

The Self-Reflective Module evaluates whether each rationale statement generated by the Generator Module is supported by the retrieved evidence and determines if further iterative refinement is necessary. Two verification mechanisms are evaluated, which are an NLI-based critic (roberta-large-mnli) and an LLM-based critic (Llama 3.1–8B). The NLI critic slightly outperforms the LLM critic, achieving a performance accuracy of 79.82% on PubMedQA and 83.33% on MedQA versus 78.76% and 82.90% respectively. The slightly better performance of the roberta-large-mnli model aligns with expectations given its entailment-focused training. Nevertheless, both critics surpass the non-critic, non-iterative baseline, demonstrating that the improve in performance is due to the iteration mechanism itself, rather than the specific critic choice.

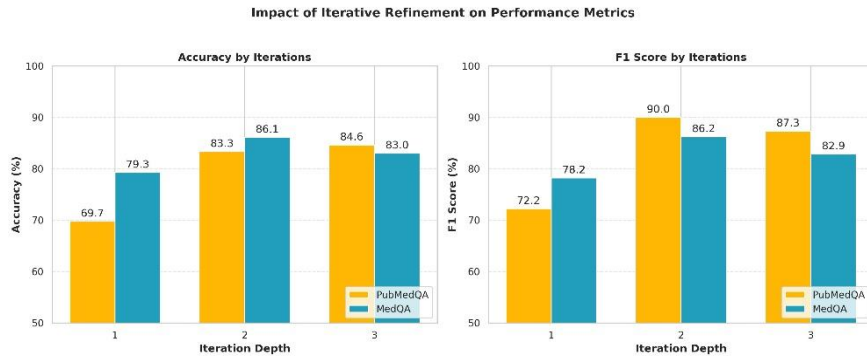

**Fig. 3. Cumulative accuracy and F1 score improvements across interations on PubMedQA and MedQA.**

Fig 3 details the cumulative impact of the iterative process done on Self-Reflective module for both accuracy and F1 scores. We observe a substantial performance leap between the first and second iterations across both datasets, with MedQA accuracy rising from 79.3% to 86.1% and PubMedQA from 69.8% to 83.3%. The upward trend confirms the potential performance gains done by the Self-Reflective module in identifying and correcting unsupported rationales. Extending the process to a third iteration, however, seems to result in a diminishing return, with performance either plateauing for PubMedQA or slightly declining for MedQA.

Overall, the findings confirm the effectiveness of Self-MedRAG's integrated design. Hybrid retrieval produces richer evidence sets, iterative reasoning improves factual grounding, and evidence based self-verification mitigates unsupported claims. These components work together to deliver robust gains across both datasets, particularly for tasks requiring multi step reasoning and synthesis across multiple biomedical documents. The results highlight the importance of retrieval diversity, iterative refinement, and rationale level verification in overcoming the limitations of traditional RAG pipelines within the biomedical domain.

## 5. Conclusions

We introduced Self-MedRAG, a self-reflective hybrid retrieval-augmented generation framework for reliable biomedical question answering. Experiments on PubMedQA and MedQA show that hybrid retrieval combining BM25 and Contriever with Reciprocal Rank Fusion substantially outperforms single retrievers, with MedQA accuracy nearly doubling compared to individual methods. Incorporating the Self-Reflective Module further improves performance, increasing PubMedQA accuracy from 69.10% to 79.82% and MedQA accuracy from 80.00% to 83.33%, demonstrating the effectiveness of iterative rationale verification in correcting unsupported claims. Both NLI-based and LLM-based critics enhance results, with the NLI critic slightly outperforming the LLM-based approach on both datasets, although the iterative refinement itself contributes the majority of the gains.

Overall, these findings validate that integrating diverse retrieval strategies with iterative rationale-level verification significantly strengthens biomedical question answering, particularly for tasks requiring multi-step reasoning and synthesis across multiple sources. They underscore the importance of retrieval diversity, self-reflective reasoning, and evidence-based verification in overcoming the limitations of traditional RAG pipelines in the biomedical domain.

**Nomenclature**

| | |
|---|---|
| $Q_0$ | Initial question/query |
| $Q_i$ | Query at iteration $i$ |
| $H$ | Reasoning history |
| $R$ | Retriever module |
| $G$ | Generator module |
| $SR$ | Self-Reflection module |
| $A$ | Final answer |
| $A_i$ | Answer at iteration $i$ |
| $Rat$ | Rationale |
| $Rat_i$ | Rationale at iteration $i$ |
| $C_i$ | Retrieved context at iteration $i$ |
| $S_i$ | Rationale support score at iteration $i$ |
| $U_i$ | Unsupported rationale elements at iteration $i$ |
| $d$ | Passage/document |
| $K$ | Reciprocal Rank Fusion constant (K = 60) |
| $i$ | Iteration index |

**Greek Symbols**

| | |
|---|---|
| $\tau$ | Verification threshold for entailment confidence ($\tau = 0.5$) |
| $\theta$ | Rationale support score threshold ($\theta = 0.7$) |

**Abbreviations**

| | |
|---|---|
| LLM | Large Language Model |
| RAG | Retrieval-Augmented Generation |
| QA | Question Answering |
| NLI | Natural Language Inference |
| RRF | Reciprocal Rank Fusion |
| BM25 | Best Matching 25 (sparse retrieval algorithm) |
| TF-IDF | Term Frequency-Inverse Document Frequency |
| FAISS | Facebook AI Similarity Search |
| USMLE | United States Medical Licensing Examination |
| MNLI | Multi-Genre Natural Language Inference |
| Acc/EM | Accuracy/Exact Match |
| F1 | F1 Score (harmonic mean of precision and recall) |
| MedCPT | Medical Contrastive Pre-trained Transformers |
| CRAG | Corrective RAG |
| SIM-RAG | Sufficiency-aware Iterative Medical RAG |
| MRD-RAG | Multi-Round Diagnostic RAG |